\documentstyle[preprint,aps]{revtex}
\begin{document}
\draft
\baselineskip 24 true pt
\vsize=9.5 true in \voffset=.2 true in
\hsize=6.0 true in \hoffset=.25 true in

\newpage
\begin{center}
{\Large{\bf A comparison of perturbative expansions using 
different phonon basis for two-site Holstein model }}\\
\end{center}
\vskip 1.0cm
\begin{center}
 Jayita Chatterjee\footnote{ e-mail: moon@cmp.saha.ernet.in} 
 and A. N. Das\footnote{ e-mail: atin@cmp.saha.ernet.in}
\end{center}
\vskip 0.50cm
\begin{center}

 {\em Saha Institute of Nuclear Physics \\
1/AF Bidhannagar, Calcutta 700064, India}\\

\end{center}

\vskip 1.0cm

\vskip 1.0cm
\begin{center}
{\bf Abstract}
\end{center}
\vskip 0.5cm
 The two-site single-polaron problem is studied within the perturbative 
expansions using different standard phonon basis obtained through 
the Lang Firsov (LF), modified LF (MLF) and modified LF transformation 
with squeezed phonon states (MLFS). The role of these convergent 
expansions using the above prescriptions in lowering 
the energy and in determining the correlation functions are 
compared for different values of coupling strength. The single-electron 
energy, oscillator wave functions and correlation functions are 
calculated for the same system. The applicability of different 
phonon basis in different regimes of the coupling strength as well as 
in different regimes of hopping are also discussed.

\newpage
\begin{center}
{\bf 1. Introduction}   
\end{center}
\vskip 0.3cm 

The Holstein model \cite {Hol} is one of the fundamental models 
describing the interactions of conduction electrons with lattice 
vibrations. The model in the simplest form consists of one electron 
hopping term, dispersionless phonons and an interaction term which 
couples the electron density and ionic displacements at a given 
site. The interaction term favours localization of the electron.  
When the interaction is strong, the gain in the localization energy 
outweighs the kinetic energy of the electron and a self-trapping of 
the electron (or a hole as the case may be) occurs with the creation of 
lattice deformation in the immediate vicinity of the   
charge carrier. The motion of the electron is then accompanied by 
the lattice deformation which results in a reduced effective hopping 
of the dressed electron (polaron). 
In the antiadiabatic limit, where the phonon frequency ($\omega_0$)  
is greater than the electronic hopping integral ($t$), the confinement 
of the lattice deformation around the charge carrier is very local 
for large electron-phonon (e-ph) interaction. This gives rise to 
small polarons whose nature and dynamics is generally studied using 
the Lang-Firsov (LF) method based on the canonical LF transformation 
\cite {LF}.  
Recently, Ranninger and Thibblin \cite {RT} and de Mello and 
Ranninger \cite{MR} studied a two-site Holstein model using the numerical 
diagonalization technique and obtained non trivial results which are 
difficult to understand from the classical zero phonon averaging LF 
approach. In fact, in Ref. \cite{MR} the authors expressed doubts 
regarding the applicability of the LF approach and validity 
of the strong coupling 
perturbation expansion (in hopping) even in the strong coupling 
antiadiabatic limit where it is believed to be valid. Considering 
a similar two-site system Firsov and Kudinov \cite{FK} studied the 
energy levels and the wave functions of the system within the 
perturbation theory 
and concluded that the results are consistent with the exact numerical 
results obtained in Refs. \cite{RT} and \cite{MR} for large e-ph 
coupling strength and small hopping. 
Marsiglio \cite {Mar} studied the Holstein model in one 
dimension with one electron up to 
16 site lattices using numerical diagonalization technique 
and concluded that neither the Migdal \cite{Mig} nor 
the usual small-polaron approximation is in quantitative agreement 
with the exact results for intermediate coupling strength. 
Kabanov and Ray \cite{KR} and Alexandrov $et~al.$ \cite {Alex} 
performed the exact numerical diagonalization of two, four and six
sites with one electron and observed that for $t> \omega_0$ the 
adiabatic Holstein small-polaron approximation, whereas for $t< \omega_0$ 
the LF approach describes the ground 
state energy of the system accurately except for a region of 
intermediate coupling strength. 
So no single conventional analytical method at present is known to 
us so as to describe the Holstein model for the entire range of the 
coupling strength either in the adiabatic or in the antiadiabatic limit. 
In a recent work \cite{DJC} we addressed this problem and considering a 
two-site system we developed a perturbation expansion using 
modified LF phonon basis. We calculated the perturbation 
corrections to the ground state energy and wave functio up to the 
third and second order respectively. The results obtained thereby 
are in good agreement with the exact numerical 
results for the entire range of the coupling strength for the 
intermediate value of hopping ($t \sim \omega_0$).   

In this paper we have extended our previous work by calculating up to 
fifth order perturbation correction to the wave function 
and sixth order correction to the energy for the ground state of 
the two-site single polaron system using various standard phonon 
basis obtained through the LF, 
modified LF (MLF) and modified LF with squeezing transformations 
(MLFS). 
It may be mentioned that the role of squeezing is believed to be 
important for intermediate coupling and hopping regime 
\cite{FCP,DS,FT,DC,CD} and studies on a two-site and a 
four-site Holstein model using the MLFS transformation followed by 
zero phonon averaging \cite {DC,CD} show that the 
energy obtained within such method is very close to the exact result. 

In the LF approach a phonon basis of fixed displacement at the 
electron residing site is chosen. Such a choice of basis diagonalizes  
the Holstein hamiltonian in absence of hopping. The hopping term 
is then treated as a perturbation \cite{MH}. The perturbation series 
is naturally expected to converge when hopping is weak and the e-ph 
coupling is strong. However, it is not precisely known the limit of 
the coupling strength as a function of hopping beyond which the 
LF approach is valid. Within the LF approach in the zeroth order 
of perturbation  
the effect of retardation between the electron and the lattice 
deformation produced by the electron cannot be obtained. 
Such a retardation becomes very important with smaller 
e-ph coupling strength and larger hopping. 
The phonon basis chosen in the MLF or MLFS approach have variational 
displacements and can produce the retardation effect even in the  
zeroth order of perturbation \cite{DS,DC}. The ground state energy 
predicted 
within the MLF and MLFS method in the zeroth order of perturbation 
is much lower than that within the LF approach \cite {CD} except for  
large values of the coupling strength, where all the methods become 
equivalent. All these suggest 
that the MLF or MLFS phonon wave functions are proper choices for 
perturbative calculation when hopping is appreciable 
and the coupling strength is not very high. For weak hopping and  
large values of the coupling strength the MLF or MLFS phonon 
wave function reduces to that of the LF approach. Thus the 
perturbation methods based on the MLF and the MLFS phonon 
wave functions have the potentiality to be applicable for a wide 
range of the coupling strength. 

The objective of this work is to develop perturbative expansions  
for the two-site single polaron system using phonon basis obtained 
through the LF, MLF and MLFS 
transformations and investigate the convergence of 
the perturbation series for different physical quantities of 
interest as a function of the e-ph coupling strength and hopping. 
Such study would provide direct answers to the important questions as  
(i) what is the region of e-ph coupling strength where the LF method  
is valid, (ii) whether the MLF and MLFS phonon wave functions are 
better than the LF wave function in the zeroth order of perturbation, 
(iii) in which region MLF and MLFS methods give better results than 
the LF method, (iv) which method within a few orders (one or two)  
of perturbation could predict results closer to the exact one. 

In Sec. II we define the model hamiltonian and describe different 
variational phonon basis states that we have considered. In sec. III 
 we present the 
expressions for the energy, wave function and static correlation 
functions calculated within the perturbation theory for the ground 
state. In Sec. IV we shall present the results obtained by 
different methods and discuss about the convergence of the 
perturbation series, hence the applicability of the methods in different  
regions of the e-ph coupling strength for different values of hopping. 

\vskip 1.0cm

\begin{center}
{\bf II. Formalism }
\end{center}
\vskip 0.3cm

The two-site single-polaron Hamiltonian is
\begin{eqnarray}
H = \sum_{i,\sigma} \epsilon n_{i \sigma} - \sum_{\sigma}
t (c_{1 \sigma}^{\dag} c_{2 \sigma} + c_{2 \sigma}^{\dag} c_{1 \sigma})
+ g \omega_0  \sum_{i,\sigma}  n_{i \sigma} (b_i + b_i^{\dag}) 
+  \omega_0 \sum_{i}  b_i^{\dag} b_i 
\end{eqnarray} 
where $i$ =1 or 2, denotes the site. $c_{i\sigma}$ ($c_{i\sigma}^{\dag}$)  
is the annihilation (creation) operator for the electron with spin 
$\sigma$ at site $i$ and $n_{i \sigma}$ (=$c_{i\sigma}^{\dag} c_{i\sigma}$) 
is the corresponding number operator, $g$ denotes the on-site e-ph coupling  
strength, $t$ is the usual hopping integral. $b_i$ and $b_{i}^{\dag}$ are the annihilation and 
creation operators, respectively, for the phonons corresponding to 
interatomic vibrations at site $i$, $\omega_0$ is the phonon frequency. 
The hamiltonian (1) has spin degeneracy, hence for the one electron case 
the spin index is redundant. So spin indices are dropped in the following.

Introducing new phonon operators $a=~(b_1+b_2)/ \sqrt 2$ and 
$d=~(b_1-b_2)/\sqrt 2 $ the Hamiltonian is separated into two parts   
($H=H_d + H_a)$ : 
\begin{eqnarray}
H_d = \sum_{i} \epsilon n_{i} - t (c_{1}^{\dag} c_{2} + c_{2}^{\dag} c_{1}) 
+ \omega_0  g_{+} (n_1-n_2) (d + d^{\dag}) 
+  \omega_0  d^{\dag} d 
\end{eqnarray}
and  
\begin{equation}
H_a =  \omega_0 \tilde{a}^{\dag}\tilde{a} - \omega_0 n^2 g_{+}^2 
\end{equation}
where $g_{+}=g/\sqrt 2$, $\tilde{a}=a +ng_{+}$ and 
$\tilde{a}^{\dag}=a^{\dag} +ng_{+}$.

$H_a$ describes a shifted oscillator which couples only  
with the total number of electrons $n(=n_1+n_2)$, which 
is a constant of motion. The last term 
in Eq.(3) represents lowering of energy achieved through 
the lattice deformations of sites 1 and 2 by the total number 
of electrons. 

$H_d$ represents an effective e-ph system where phonons directly 
couple with the electronic degrees of freedom and its solution by 
any analytical method is a non trivial problem.   
We now use the MLF transformation where the lattice deformations 
produced by the electron are treated as variational parameters 
\cite {DS,DC,LS}. For the present system, 
\begin{equation}
\tilde{H_d} = e^R H_d e^{-R}
\end{equation}
where $R =\lambda (n_1-n_2) ( d^{\dag}-d)$, $\lambda$ is a
variational parameter related to the displacement of the $d$ 
oscillator. 

The transformed Hamiltonian is then obtained as 
\begin{eqnarray}
\tilde{H_d}&=&\omega_0  d^{\dag} d + \sum_{i} \epsilon_p n_{i} - 
t [c_{1}^{\dag} c_{2}~ \rm{exp}(2 \lambda (d^{\dag}-d))    \nonumber\\ 
&+& c_{2}^{\dag} c_{1}~\rm{exp}(-2 \lambda (d^{\dag}-d))]  
+ \omega_0  (g_{+} -\lambda) (n_1-n_2) (d + d^{\dag}) 
\end{eqnarray}
where 
\begin{equation}
\epsilon_p = \epsilon - \omega_0 ( 2 g_{+} - \lambda) \lambda 
\end{equation}

 For a perturbation method it is desirable to use a basis 
where the major part of the hamiltonian becomes diagonal. 
When the hopping is 
appreciable a retardation between the electron and associated 
lattice distortion sets in and as mentioned before the MLF or 
the MLFS method would work better than the LF method for a wide 
region of e-ph coupling strength.

Now we will make a squeezing transformation \cite{HZ} to the 
Hamiltonian (5)  
\begin{eqnarray} 
\tilde{H_{sd}} &=& e^S \tilde{H_{d}} e^{-S}   
\end{eqnarray} 
where $S=\alpha(d_id_i-d^{\dag}_{i} d^{\dag}_{i})$. The new phonon 
basis is squeezed with respect to the previous basis. 
Squeezing parameter ($\alpha$) partly reduces 
the polaronic narrowing effect and consequently enhances the hopping.
However, the phonon energy increases with increasing $\alpha$ and a 
competition between phonon energy and hopping delocalization (kinetic)
energy determines the value of $\alpha$ \cite{DKR}.
The transformed hamiltonian (7) takes the form 
\begin{eqnarray} 
\tilde{H_{sd}} &=& \omega_0  d^{\dag} d [\rm{cosh}^2(2\alpha)
 +\rm{sinh}^2(2\alpha)] + \omega_0~\rm{cosh}(2\alpha)~\rm{sinh}(2\alpha)  
(dd+ d^{\dag}d^{\dag})  \nonumber\\  
&+& \sum_{i} \epsilon_p n_{i} 
- t [c_{1}^{\dag} c_{2}~\rm{exp}(2 \lambda_{e} (d^{\dag}-d))    
+ c_{2}^{\dag} c_{1}~\rm{exp}(-2 \lambda_{e} (d^{\dag}-d))] \nonumber\\  
&+& \omega_0  (g_{+} -\lambda) (n_1-n_2) (d + d^{\dag})~\rm{exp}(2 \alpha)
+\omega_0~\rm{sinh}^2(2\alpha)  
\end{eqnarray} 
\begin{equation}
\rm{where}~~\lambda_{e}=  \lambda ~e^{-2\alpha} \nonumber
\end{equation}

For the single polaron problem we choose the basis set 
\begin{equation}
|\pm,N \rangle = \frac{1}{\sqrt 2} (c_{1}^{\dag} \pm  c_{2}^{\dag}) 
|0\rangle_e  |N\rangle 
\end{equation}
where $|+\rangle$ and $|-\rangle$ are the bonding and antibonding 
electronic states and $|N\rangle$ denotes the $N$th excited oscillator  
state in the MLFS, MLF or LF basis depending on the method considered. 
It may be noted that the MLFS basis turns into the MLF basis if one 
puts $\alpha =0$ and the MLF into the LF basis when $\lambda= g_+$. 

The hopping term $H_t$ in Eq. (8) has both diagonal and nondiagonal 
matrix elements in the chosen basis (10). The 
diagonal part of $H_t$ is given by 
\begin{eqnarray}  
\langle N, \pm|H_t|\pm , N\rangle &=& \langle N, \pm|- t [c_{1}^{\dag} 
c_{2}~ \rm{exp}(2 \lambda_{e} (d^{\dag}-d)) + c_{2}^{\dag} 
c_{1}~\rm{exp}(-2 \lambda_{e} (d^{\dag}-d))]|\pm,N\rangle \nonumber\\
&=& \mp t_{e}\sum_{i=0}^{N}\left[
\frac{(2\lambda_{e})^{2i}}{i!}(-1)^i N_{C_i}\right]
\end{eqnarray}  
where $t_{e}=t~\rm{ exp}{(-2\lambda_{e}^2)}$ and 
$N_{C_i}=\frac{N!}{i!(N-i)!}$.
              
For the perturbation method we consider the diagonal part of the 
Hamiltonian (8) as the unperturbed Hamiltonian ($H_0$) and the 
remaining part of the Hamiltonian, $H_{1}= \tilde{H_{sd}}-H_0$,  
as a perturbation. 

The unperturbed energy of the state $| \pm,N\rangle$ is given by 
\begin{eqnarray}
 E_{\pm,N}^{(0)}= \langle N,\pm|H_0|\pm, N \rangle=
 \omega_0 [\rm{sinh}^2(2\alpha) + N(\rm{sinh}^2(2\alpha) 
 + \rm{cosh}^2(2\alpha))]\nonumber\\
 + \epsilon_p \mp t_{e} \left[ \sum_{i=0}^{N}
 (\frac{(2\lambda_{e})^{2i}}{i!} (-1)^i N_{C_i}\right]
\end{eqnarray}
The general off-diagonal matrix elements of 
$H_1$ between the two states $|\pm,N \rangle$ and $|\pm,M \rangle$ 
are calculated as (for $(N-M)>0$)
\begin{eqnarray}  
\langle N,\pm|H_1|\pm, M \rangle&=&P(N,M)+ 
\frac{\omega_0}{2}\sqrt{N(N-1)}~\rm{sinh}(4\alpha) 
~\delta_{N,M+2} \\
&&~~~~~~~~~~~~~~~~~~~~\rm{for}~\rm{even}~(N-M).\nonumber \\ 
 \nonumber \\
\langle N,\pm|H_1|\mp,M \rangle&=&P(N,M)+ \sqrt{N}\omega_0 
(g_{+}-\lambda) e^{2\alpha} \delta_{N,M+1}\\
&&~~~~~~~~~~~~~~~~~~~\rm{for}~\rm{odd}~ (N-M). \nonumber\\
\rm{where}~~~ P(N,M)&=&\mp t_{e} (2\lambda_{e})^{N-M} 
\sqrt{\frac{N!}{M!}} \left[  \frac{1}{(N-M)!}+\sum_{R=1}^{M}
[(-1)^R \right. \nonumber \\
&& \left. \frac{(2\lambda_{e})^
{2R}}{(N-M+R)! R!}M(M-1)...(M-R+1) ] \right] \nonumber
\end{eqnarray}  

\vskip 1.0cm
\begin{center}
{\bf III. Perturbation corrections to the ground state}
\end{center}
\vskip 0.3cm

In this section we present the calculations of the perturbation 
corrections to the ground-state energy and wave function up to 
the sixth and fifth order, respectively using the LF, MLF and 
MLFS methods. The static correlation 
functions relating the electron and associated lattice deformations 
are also calculated using the corresponding 
perturbed wavefuctions.

For the chosen basis (10) the state $|+\rangle|0\rangle$ has the 
lowest unperturbed energy, $ E_0^{(0)}=\epsilon_p-t_{e}+ \omega_0 
~\rm{sinh}^2(2\alpha)$. The matrix element connecting this 
ground state and an excited state $|e,N\rangle$ is given by
\begin{eqnarray}  
\langle N,e|H_{1}|+,0\rangle&=&\left[ -t_{e}\frac{(2\lambda_{e})^N}
{\sqrt{N!}}+\omega_0(g_{+}-\lambda)e^{2\alpha} ~\delta_{N,1} \right] 
\delta_{e,-}\hspace{.2cm} \rm{for~odd ~ N } \\
&=& \left[ -t_{e}\frac{(2\lambda_{e})^N}{\sqrt{N!}} 
+\sqrt{2}\omega_0~\rm{sinh}(2\alpha)~\rm{cosh}(2\alpha)~
\delta_{N,2}\right]
 \delta_{e,+} ~\rm{for~ even~ N }\nonumber 
\end{eqnarray}  
 
The first order correction to the ground state wave function is 
obtained as,  
\begin{eqnarray}  
|\psi_0^{(1)}\rangle &=&\frac{[\omega_0(g_{+}-\lambda)e^{2\alpha}-
2\lambda_{e} t_{e}]}{(E_{0}^{(0)}-E_{+,1}^{(0)})}~|-,1\rangle 
+\frac{[-t_{e}(2\lambda_{e})^2+2\omega_0~ 
\rm{sinh}(2\alpha)~\rm{cosh}(2\alpha)]}{\sqrt{2!}(E_0^{(0)}-E_{+,2}^{(0)})}~
|+,2\rangle \nonumber\\
&-& \sum_{N=3,4,5..}\frac{t_{e}(2\lambda_{e})^N}{\sqrt{N!}
( E_0^{(0)}-E_{e,N}^{(0)})}~|e,N\rangle 
\end{eqnarray}  
where $E_{\pm,N}^{(0)}$ is the unperturbed energy of the state 
$|\pm,N\rangle$ as given in Eq.(12) and e = + or - for even and odd N 
respectively.

The first order correction to the energy ($E_0^{(1)}$) is zero since 
$H_{1}$ has no diagonal matrix 
element. The second order correction to the ground state energy 
is given by   
\begin{eqnarray}
E_0^{(2)} &=& +\sum_{N=1,3..}\frac{|\frac{-t_{e}(2\lambda_{e})^{N}}
{\sqrt{N!}}+\omega_0 (g_+ - \lambda)~e^{2\alpha}~\delta_{N,1}|^2}
{(E_0^{(0)}-E_{-,N}^{(0)})}\nonumber\\ 
&+&\sum_{N=2,4..}\frac{|\frac{-t_{e}(2\lambda_{e})^{N}}
{\sqrt{N!}}+ \sqrt{2}~\omega_0~\rm{sinh}(2\alpha)~\rm{cosh}(2\alpha)
~\delta_{N,2}|^2}
{(E_0^{(0)}-E_{+,N}^{(0)})} 
\end{eqnarray}
Now one has to make a proper choice of $\lambda$ and $\alpha$ so 
that the perturbative expansion becomes convergent. Usually,  
within the MLF and MLFS approach zero phonon averaging is made 
\cite{DS,DC} and the variational parameters $\lambda$ and $\alpha$ are 
found out by minimizing the ground state energy of the system. 
This corresponds to the minimization of the unperturbed energy in 
our calculation. Previously we have followed this procedure 
within the MLF approach \cite{DJC} and found that the perturbation 
corrections to the energy converge rapidly.
Here we will follow the same procedure to find out the variational  
phonon basis as a function of e-ph coupling for the MLF and MLFS 
methods. Minimizing the unperturbed ground state energy $E_0^{(0)}$ 
with respect to $\lambda$ and $\alpha$ we obtain 
\begin{eqnarray}
\lambda=\frac {\omega_0g_{+}}{\omega_0+2t_{e} e^{-4\alpha}} \nonumber\\ 
\alpha = \frac{1}{4}\rm{sinh}^{-1}[2\alpha(g_+-\lambda)]    \nonumber 
\end{eqnarray}
It is interesting to note that these choices of $\lambda$ and $\alpha$ 
make the offdiagonal matrix elements between the states 
$|+,0\rangle$ to $|-,1\rangle$ and $|+,2\rangle$ equal to zero and 
consequently, the coefficients of $|-,1\rangle$ and $|+,2\rangle$ in 
Eq.(16) and the terms corresponding to $N=1,2$ in the r.h.s of 
Eq.(17) vanish. 
Since the first two terms in Eq. (16) are zero for the MLFS choice of 
the basis it is expected that the first order correction to the 
ground state wave function and second order correction (Eq. 17) to 
the energy would be smaller within the MLFS method. 
To examine the validity of the perturbation procedure one should 
calculate the higher order corrections and check whether the series 
is converging properly. 

The second order corrections to the ground state 
wave function is given by 
\begin{eqnarray}  
|\psi_0^{(2)}\rangle = \sum_{k \neq 0} a_{\rm{0k}}^{(2)} |\rm{k}\rangle \\
 \rm{where} ~~~
a_{\rm{k0}}^{(2)} = \sum_{m\neq 0}[ \frac{(H_1)_{km}
(H_1)_{m0}}{(E_{0}^{(0)}-E_k^{(0)})(E_{0}^{(0)}-E_m^{(0)})}]\nonumber
\end{eqnarray}  
where $|\rm{k}\rangle$'s denote the unperturbed states 
$|\pm,N\rangle$ with the unperturbed energy $E_{\pm,N}^{(0)}$ and 
the 0 refers to the ground state $|+,0\rangle$. $(H_1)_{\rm{km}}$ is the 
off-diagonal matrix element of $H_1$ between the states $|\rm{k}\rangle$
and $|\rm{m}\rangle$. These matrix elements are given in Eqs. (13) 
and (14). 
The third order correction ($ E_0^{(3)}$) to the ground state 
energy is given by,
\begin{eqnarray}
E_0^{(3)} &=& \sum_{k\neq 0} (H_1)_{\rm{0k}} a_{\rm{0k}}^{(2)}
\end{eqnarray}

In general the nth order correction to the wave function and the 
(n+1)th order correction to the energy are given by 
\begin{eqnarray}
|\psi_0^{(n)}\rangle = \sum_{k\neq 0} a_{\rm{0k}}^{(n)} |\rm{k}\rangle \\
E_0^{(n+1)} = \sum_{k\neq 0} (H_1)_{0k} a_{\rm{0k}}^{(n)}
\end{eqnarray}
where  
\begin{eqnarray}
a_{\rm{0k}}^{(n)}&=& \frac{1}{(E_{0}^{(0)}-E_k^{(0)}) } 
\left[- E_0^{(1)}a_{\rm{0k}}^{(n-1)}-E_0^{(2)}a_{\rm{0k}}^{(n-2)} \right.
- E_0^{(3)}a_{\rm{0k}}^{(n-3)}-......-E_0^{(n-1)}a_{\rm{0k}}
^{(1)} \nonumber\\
&+&\left.\sum_{m\neq 0}[ (H_1)_{\rm{km}} a_{\rm{0m}}^{(n-1)}]~\right]
\end{eqnarray}
Using Eqs. (20) and (21) all the higher order corrections to the 
wave function and energy may be calculated step by step.  

The ground state wave function may be written as,
\begin{eqnarray}  
|\psi_{G} \rangle & \equiv& |+,0\rangle+|\psi_0^{(1)}\rangle + 
|\psi_0^{(2)}\rangle+ |\psi_0^{(3)}\rangle + ......  \nonumber\\
 &=& |+,0\rangle 
 +\sum_{N=1,3,..}c_{N} |-,N \rangle
 +\sum_{N=2,4,..}c_{N}|+,N \rangle  
\end{eqnarray} \\
The coefficients $c_{N}$ are determined from the 
sum of the corresponding coefficients $a_{\rm{0k}}^{(1)}$, 
$a_{\rm{0k}}^{(2)}$, $a_{\rm{0k}}^{(3)}$, etc., 
where $N$ represents the state $|\rm{k}\rangle$. 
The normalized ground state wave function $|\psi_{G} \rangle_{N}$ is 
$$|\psi_{G} \rangle_{N}=\frac{1}{\sqrt{N_G}}|\psi_{G} \rangle$$ 
where $N_G$ is obtained as
\begin{eqnarray}  
N_G \equiv \langle \psi_{G}|\psi_{G}\rangle =1
+\sum_{N=1,2,3,..}c_{N}^2\nonumber 
\end{eqnarray} 
\vskip .5cm
\begin{itemize}
\item {Correlation function}
\end{itemize}

   The static correlation functions $\langle n_1 u_{1}\rangle_{0}$ and 
$\langle n_1 u_{2}\rangle_{0}$,
where $u_1$ and $u_2$ are the lattice deformations at sites 1 and 2  
respectively, produced by an electron at site 1, are the standard 
measure of polaronic character and indicate the strength of polaron 
induced lattice deformations and their spread. The operators 
involving the correlation functions may be written as 
\begin{equation}  
n_{1} u_{1,2} = \frac{n_1}{2}[(a+a^{\dag})\pm(d+d^{\dag})
e^{2\alpha} -~2~(n g_+\pm \lambda(n_1-n_2))]\nonumber \\
\end{equation}
The final form of the correlation functions are obtained as
\begin{eqnarray}
\langle n_{1} u_{1} \rangle_{0}&=&\frac{1}{2}  
\left[-(g_{+} +\lambda) + \frac{A_0~e^{2\alpha}}{N_G}\right] \\
\langle n_{1} u_{2} \rangle_{0}&=&\frac{1}{2}
\left[-(g_{+}-\lambda)- \frac{A_0~e^{2\alpha}}{N_G}\right] \nonumber
\end{eqnarray}  
\begin{eqnarray} 
\rm{where}~~~A_0\equiv\langle \psi_{G} |n_1(d+d^{\dag})|\psi_{G}
\rangle=\sum_{N=1,3,5..}c_N\left[
\sqrt{N}~c_{N-1}+\sqrt{N+1}~c_{N+1}\right] \nonumber
\end{eqnarray} 
\vskip 1.0cm
\begin{center}
{\bf IV. Results and discussions}
\end{center}
\vskip 0.3cm

  In this paper we present the results for $t$=0.5, 1.1  and 
2.1 (in a scale of $\omega_0$ =1). These values of $(t/\omega_0)$   
covers the cross-over region from the antiadiabatic to the adiabatic 
limits. Furthermore, exact results on some of the physical quantities 
are available for the above values of $(t/\omega_0)$ which enables 
us to compare our results obtained within the perturbation methods with 
the exact results. 

We estimate the perturbation corrections to the ground state energy 
and wave function up to the sixth and fifth orders respectively, 
within the LF, MLF and MLFS methods considering 25 phonon states 
(which is sufficient for $g_+\leq ~2.2$) in the transformed 
phonon basis.   
Calculations of such successive orders of corrections would 
provide us the key features of the convergence of the perturbation 
series and the applicability of the method concerned. 

For $(t/\omega_0)=0.5$ we find a rapid convergence in the perturbation 
series for the ground state energy for all values of $g_+$. Fig. 1   
shows the relative perturbation corrections i.e., the ratios of 
the perturbation corrections (to the ground state 
energy) of different orders to the unperturbed energy as a function 
of $g_+$. The 
energy corrections within the MLFS and the MLF methods are much  
smaller than that within the LF method for $g_+ <1.6$. For 
$(t/\omega_0)=0.5$ the fifth or sixth order correction to the 
energy is so small that the energy obtained with fifth or sixth 
order of perturbation within MLF method may be treated as the 
exact energy for all 
values of $g_+$. In Fig.2(a) we plot the exact energy thus  
obtained along with the unperturbed energies within different 
methods. It is seen that the MLFS or the MLF unperturbed energies are 
closer to the exact one than the LF unperturbed energy. In 
Fig. 2(b) we show the variation of 
$|\langle \psi_0^{(0)}|\psi_G\rangle|^2$, which gives the probability 
that the perturbed ground state ($|\psi_G\rangle$) lies in the 
unperturbed ground state $|\psi_0^{(0)}\rangle$, with $g_+$ 
($|\psi_G\rangle$ is found out considering up to the fifth order 
correction to the wave function). This probability is closer 
to 1 for the MLFS and the MLF methods compared to that in the LF 
method. The results of Fig. 2 clearly indicate that the phonon basis, 
chosen within the MLFS or the MLF methods, are much better than the  
LF basis. It may be mentioned that the  
correlation function $\langle n_{1} u_{2} \rangle_{0}$, which 
is a measure of the lattice deformation at site 2 
produced by an electron at site 1 due to retardation effect vanishes 
for the LF unperturbed state. This correlation function is very 
sensitive to the higher order 
perturbation corrections to the wave function
 than the ground state energy. 
 To examine the convergence of the perturbative expansion on 
this correlation function we plot $\langle n_{1} u_{2} \rangle_{0}$,   
calculated  up to different orders of perturbation corrections, 
as a function of $g_+$ in Fig. 3. It is found that for 
$(t/\omega_0)=0.5$ this correlation function converges for all values 
of $g_+$. The curves corresponding to the correlation function 
calculated up to 
fourth and fifth order corrections to the wave function almost merge 
together (Fig.3) and specially for the MLF method they are almost 
identical in the entire region of $g_+$. Hence, the correlation 
function obtained considering up to the fifth order correction within  
the MLF method, may be considered as the exact one for $(t/\omega_0)=0.5$.  
Fig. 3 also shows that even in the strong coupling limit one has to 
consider up to the second order corrections to the wave function to obtain 
(nearly) exact results. It is interesting to note that 
the MLF and the MLFS methods can 
predict non zero values of $\langle n_{1} u_{2} \rangle_{0}$ even 
in the zeroth order of perturbation (i.e. with the unperturbed ground 
state wave function in the MLF or MLFS basis) for low and intermediate 
values of $g_+$ whereas the unperturbed LF wave function 
always predicts zero value for 
$\langle n_{1} u_{2} \rangle_{0}$ and we find that in an appreciable 
region of low values of $g_+$  the predictions of the MLFS method 
with zeroth order of perturbation is very close to the exact one. 

In Figs. 4-8 we have shown the results for $t/\omega_0=1.1$ which 
is in between the adiabatic and antiadiabatic limits. 
In Fig. 4 the 
relative perturbation corrections to the unperturbed energy are plotted 
as a function of $g_+$. The perturbation corrections in energy 
are smaller within the MLF and MLFS than the LF method. For each 
method there is a range of $g_+$ where higher order corrections are 
appreciable and the convergence is weak. For $t/\omega_0=1.1$ 
the range of $g_+$ where the convergence is weaker, 
is 0.4$-$1.2 for the LF, 1$-$1.3 
for the MLF and 1.1$-$1.55 for the MLFS method. 
For values of $g_+$ outside this range 
the perturbation corrections show rapid convergence. In the 
MLF method the perturbation corrections show satisfactory 
convergence in the entire region of $g_+$, while for the LF and 
MLFS methods there are small regions of $g_+$ where 
fifth or sixth order corrections are comparable or 
greater than the third and fourth order corrections and 
the convergence is not satisfactory in that region. 
The ground state energy calculated within the MLF method considering 
up to the sixth order correction, is shown in Fig. 5 as a function 
of $g_+$. This energy when compared with the exact results of  
Ref. \cite{RT}, is found to be almost identical with the exact one.  
This is expected because the higher order corrections are  
successively smaller in magnitude and of alternate sign within the 
MLF method, so the net contributions from seventh to higher 
order terms to the energy would be negligibly small with respect 
to the unperturbed energy. 
The unperturbed ground state energy for the LF, MLF and the MLFS 
methods are also shown in the same figure. 
The MLFS unperturbed energy is closer to the exact energy and for low 
values of $g_+$ it is almost identical to the exact energy.  

In Fig. 6 we plot the correlation function 
$\langle n_{1} u_{2} \rangle_{0}$ obtained by considering up to the 
different orders of perturbation corrections to the wave function 
against $g_+$. The MLFS method shows excellent convergence for low 
values 
of $g_+$ ($\le 0.9$), the LF method shows very good convergence 
beyond $g_+=1.2$. The MLF method shows good convergence for 
the entire region of $g_+$, however the convergence is weaker in the 
range $0.9 \le g_+ < 1.3$. When compared with exact results of 
$\langle n_{1} u_{2} \rangle_{0}$ (taken from the Ref.\cite{RT}), 
it is found that the MLF results up to the fifth order perturbation 
are identical to the exact results except in the region 
$0.9 < g_+<1.3$ where a slight departure in values from the exact 
results is seen. The correlation function $\langle n_{1} u_{1} \rangle_{0}$
would evidently show a much more rapider convergence than the 
convergence of $\langle n_{1} u_{2} \rangle_{0}$. 

To examine the nature of the cross-over from the delocalized (large) to 
localized (small) polaron we plot 
$\langle n_{1} (u_{1}-u_{2}) \rangle_{0}/g_+$
against $g_+$ in Fig. 7 where the (nearly) exact plot (obtained  
from the MLF considering up to fifth order corrections to the 
wave function) and those obtained from the MLF and MLFS in the 
zeroth order of perturbations are shown. 
The standard LF approximation (zeroth order of perturbation  
or zero phonon averaging) always predicts the value of this quantity 
to be equal to 1, which is a characteristic of extremely localized 
polarons, and so a cross-over from small to large 
polaron cannot be obtained in the standard LF approximation. The MLF 
and MLFS methods can predict this cross-over even in the zeroth 
order of perturbation \cite{DS,DC,CD}. It may be mentioned  
that the value of $\langle n_{1} (u_{1}-u_{2}) \rangle_{0}/g_+$  
within MLF and MLFS methods in the zeroth order of perturbation is 
$\lambda /g_+$ (Eq. (25)). Fig. 7 shows that up to 
$g_+=1.0$ the results of MLFS with zero phonon averaging almost 
coincides with the exact results. The (nearly) exact plot shows  
a smooth cross-over from the delocalized to localized 
polaron with increase of $g_+$, whereas the zero phonon 
averaging results within MLF or MLFS procedure shows an abrupt 
feature which were noted before \cite{DS,DC,CD}. L\.owen 
\cite{L} pointed out that for a finite phonon frequency there 
cannot be any abrupt cross-over in the ground state of an e-ph system. 
The (nearly) exact plot of Fig. 7 is consistent with this conclusion. 

In Fig. 8 we have shown the ground state wave function (within the MLF 
method considering up to fifth order correction to the 
wave function) for the $d$ 
oscillator as a function of position $x$ for different values of 
the e-ph coupling when the electron is located at site 1. 
For weak coupling the wave function shows displaced  
Gaussian like single peak. 
For intermediate values of $g_+$ an additional prominent shoulder 
appears in the wave function as seen in the curve for $g_{+}$=1.3. 
For higher values of $g_{+}$ this shoulder takes the form of a broad 
peak. These results are completely consistent 
with the results obtained by Ranninger and Thibblin by exact 
diagonalization study \cite{RT}. 

In Figs. 9-12 we have given the results of $t/\omega_0 =2.1$,   
which is towards the adiabatic limit where the LF method 
is not expected to work in this region. So it 
would be of general interest to examine whether the methods, 
dealt in this paper, show good convergence in any region of $g_+$.   
In Fig. 9 the relative perturbation corrections of different orders to 
the ground state energy are shown as a function of $g_+$.  
The figure shows that the perturbation corrections within the LF 
method are very large and do not converge at all in an appreciable 
region of $g_+$ (0.6-1.25). For the MLF and MLFS methods the 
perturbation corrections in energy are much smaller. The MLF method 
shows good convergence except in a region of $1.2 <g_+< 1.6$. 
For the MLFS method the 
corrections are negligible in a wide region of $g_+$ ($0 <g_+ <1.1$). 
However, in a region 
$1.25 <g_+< 1.7$ the perturbation corrections within the MLFS do 
not converge. For $g_+>1.4$ the convergence within the LF method 
is satisfactory and hence may be used. 

In Fig. 10 we show the variation of 
$|\langle \psi_0^{(0)}|\psi_G\rangle|^2$ with $g_+$ for different methods. 
The figure shows that for $t/\omega_0=2.1$ this value within the LF 
method is far from 1.0 for a broad region of $g_+$ (0.4$-$1.3) which 
signifies that the LF method is totally inapplicable in that 
region. However, in the same region the value of 
$|\langle \psi_0^{(0)}|\psi_G\rangle|^2$ within the MLFS is close to 
1 and thus MLFS should work exceedingly well there.  

In Fig. 11 the convergence of the correlation function 
$\langle n_{1} u_{2} \rangle_{0}$ is shown. It is clearly 
found that the MLFS 
method shows excellent convergence for low values 
of $g_+$ ($\le 1.2$), the LF method shows good convergence 
beyond $g_+=1.5$. The MLF method shows a systematic good 
convergence for the entire region of $g_+$, however the convergence is 
weaker in an intermediate region of $g_+$. 

In Fig. 12  the ground state phonon wave functions, evaluated 
by considering up to the fifth order perturbation correction within the 
MLF and LF methods, are shown for the $d$ oscillator as a function of 
position $x$ for $g_+= 1.6$ and for different values of $t/\omega_0$ 
(0.5, 1.1, 2.1), when the electron is located at site 1. 
The LF and the MLF predictions are identical and hence cannot be 
distinguished in the plot. These plots also exactly match with 
the phonon wave functions, calculated with exact  
diagonalization technique \cite{MR}, for the corresponding set of 
parameters. The reason for such excellent agreement with 
the exact results is that for $g_+=1.6$ and for 
$t/\omega_0= 0.5,~ 1.1~\rm{and}~ 2.1$, a very good convergence 
is achieved in the wave function within the LF 
and MLF methods which is evident from Figs. (3), (6) and (11). 

\vskip 1.0cm
\begin{center}
{\bf V. Conclusions}
\end{center}
\vskip 0.3cm

In the present work we examine the convergence of the perturbation 
expansions using the LF, MLF and MLFS phonon basis for a two-site 
one electron Holstein model for three values of 
$t/\omega_0~ (0.5,~ 1.1~\rm{and}~ 2.1)$. For $t/\omega_0=0.5$ 
the perturbation 
series show very good convergence within all the three methods 
in the entire region of e-ph coupling strength. 
The perturbation corrections within the MLF and MLFS methods 
are much smaller than that of the LF method. For higher values of 
$t/\omega_0$ also good convergence is achieved for the major region of 
$g_+$. The perturbation expansion shows weak or bad convergence in 
an intermediate region of $g_+$, the range of which varies in 
different methods.   
The MLFS method yields negligible perturbation corrections and 
shows good convergence for a wide region of $g_+$ (from zero to 
an intermediate value) which increases with increasing value 
of $t/\omega_0$. In this region the LF method particularly shows 
bad convergence hence inapplicable. 
For large values of $g_+$ the LF method shows good convergence, as 
expected; the MLF and the MLFS methods also work satisfactorily there. 
The MLF method shows a systematic convergence in the whole region 
of $g_+$ even for higher values of $t/\omega_0$, as considered here. 
It is interesting to note that for strong coupling strength 
consideration of the second order correction to the energy as well 
as to the wave function reproduces almost the exact results for 
all the three methods. 
The calculated oscillator wave functions following our method 
match excellently with the exact oscillator wave functions, derived from 
the numerical diagonalization 
studies. Our study also shows that the region of $g_+$, where no  
perturbation analytical method is applicable, is really very narrow 
in contrast to the conclusions of some earlier works  
\cite {RT,MR}. Although the analysis presented in this paper 
is based on a two-site system, the approach may permit 
a better progress for analytical studies of many-site systems.

\newpage
%\begin{references}

\newpage
Figure captions :

\noindent
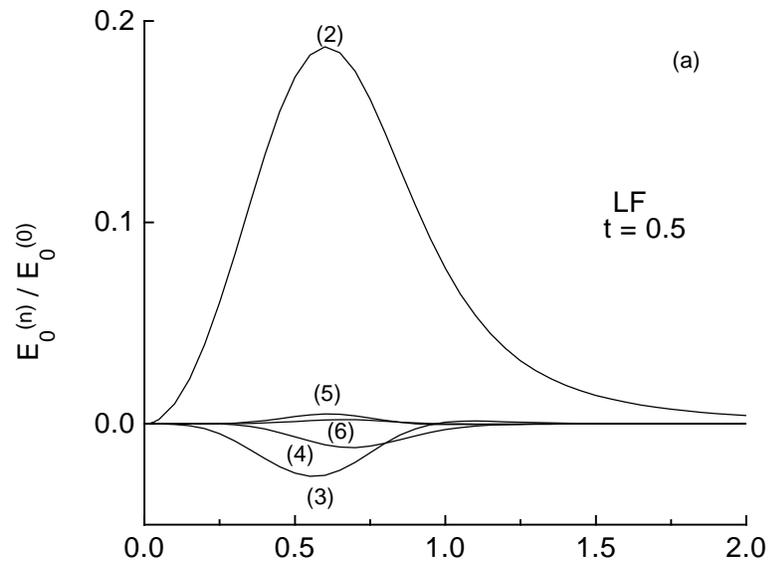
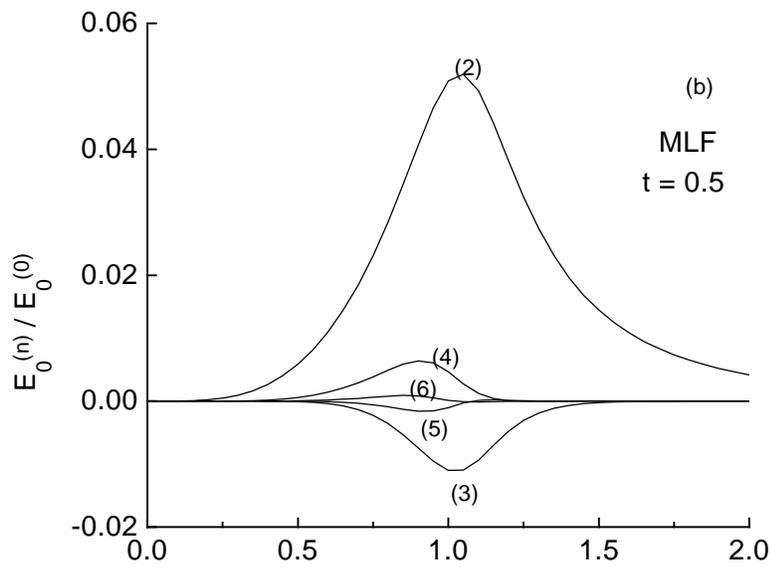
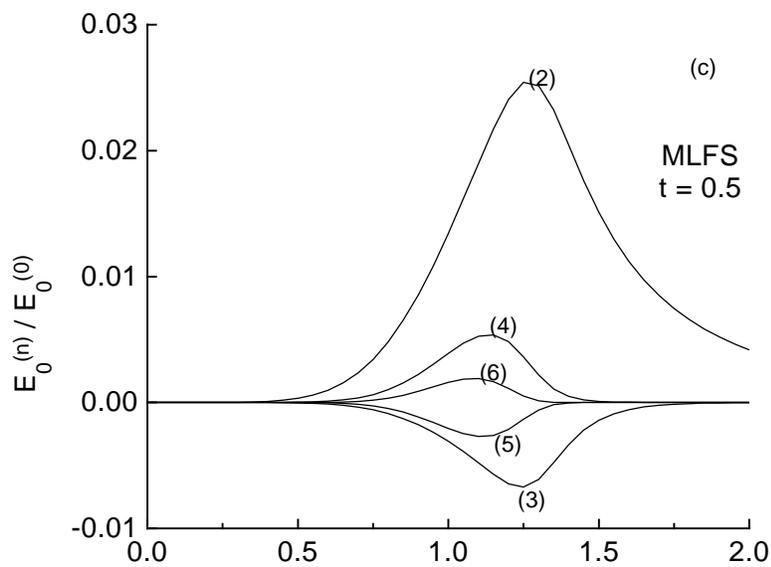
FIG. 1. Variation of the relative perturbation corrections 
$E_{0}^{(n)}/E_{0}^{(0)}$ to the ground state energy as a 
function of the coupling strength ($g_{+}$) for $t/\omega_0 =0.5$ 
for (a) LF, (b) MLF and (c) MLFS method. 
$E_{0}^{(n)}$ is the nth order perturbation correction to the 
ground state energy and 
$E_{0}^{(0)}$ is the unperturbed ground state energy. 

\vskip 0.5 cm
\noindent

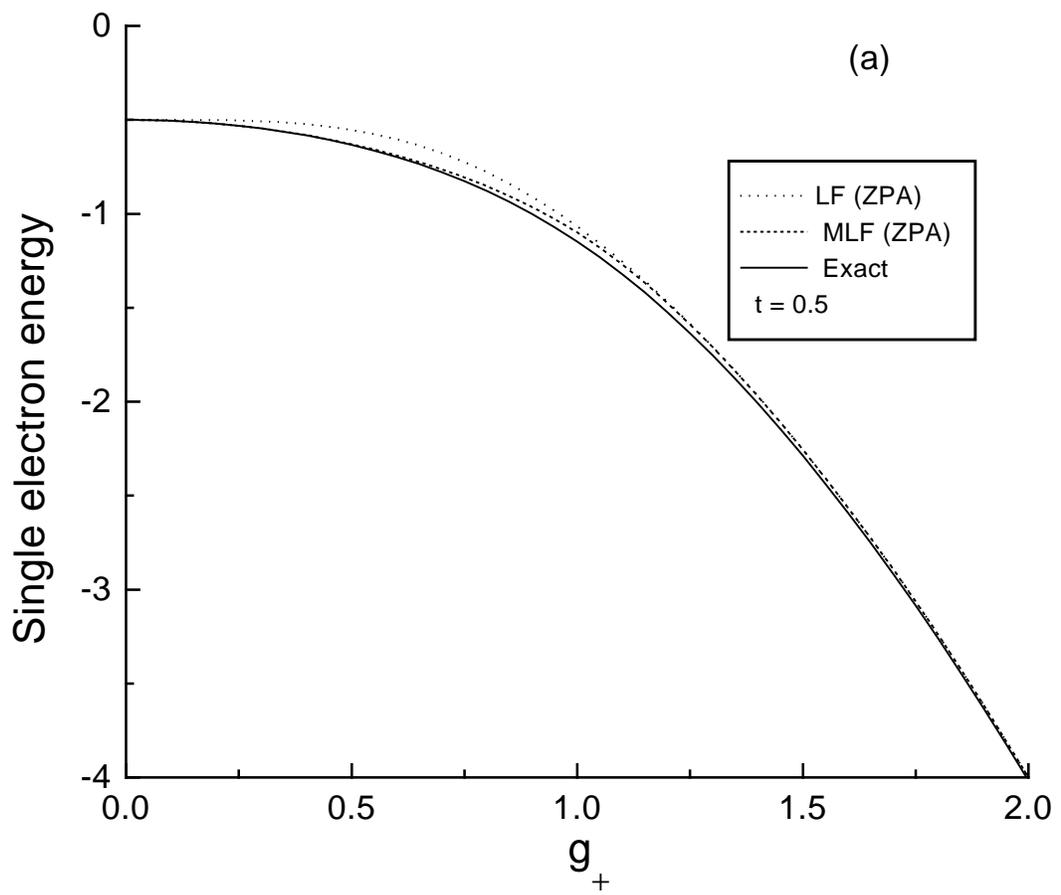
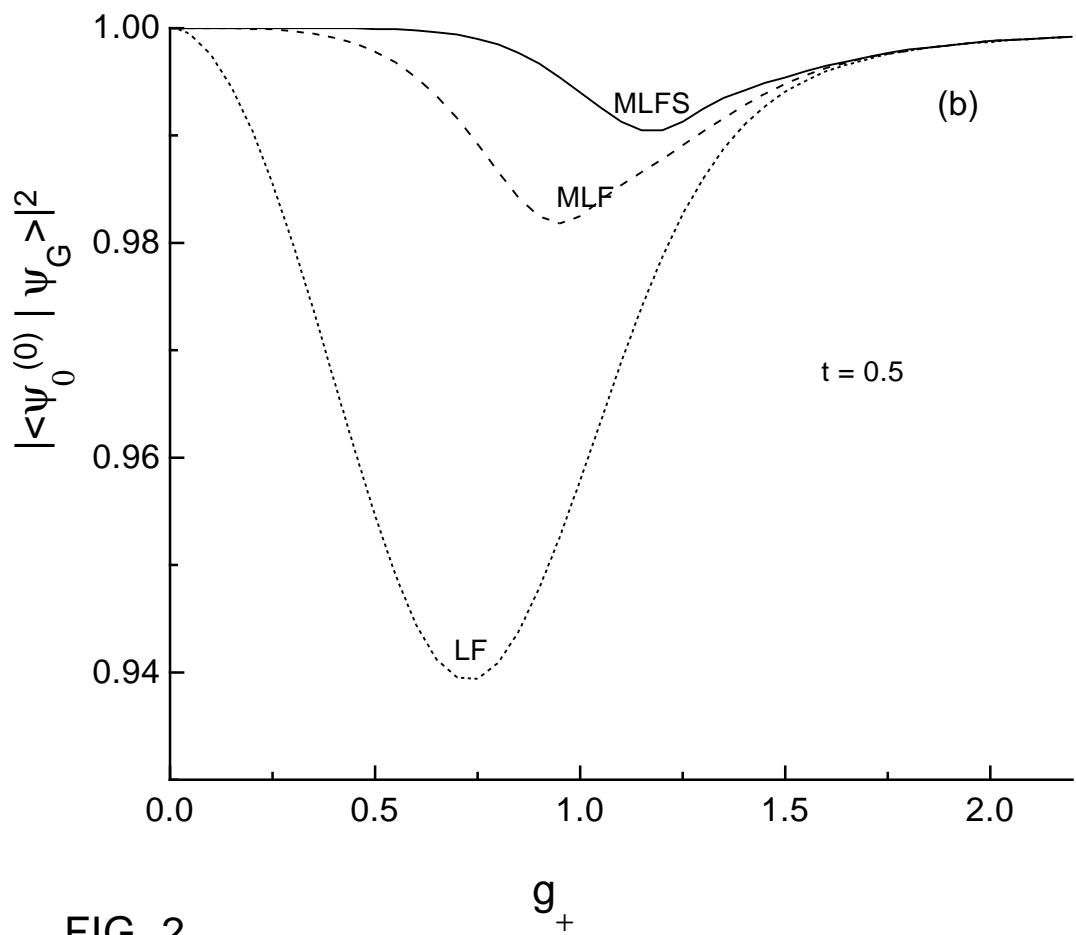
FIG. 2.(a): The plot of the ground state energy obtained by considering 
perturbation correction up to fifth order within the MLF method as 
a function of $g_+$ for $t/\omega_0 =0.5$. 
The unperturbed energies (equivalent to the energy obtained with 
standard zero phonon averaging (ZPA)) within the LF and MLF methods are 
also shown. The MLFS (ZPA) results (not shown) 
lies in between the MLF and the exact curves.  
(b): The variation of $|\langle \psi_0^{(0)}|\psi_G\rangle|^2$ with 
 $g_+$ within LF, MLF and MLFS methods. $|\psi_G\rangle$ and 
$|\psi_0^{(0)}\rangle$ are the corrected and the unperturbed wave 
functions, respectively.
\vskip 0.5 cm
\noindent

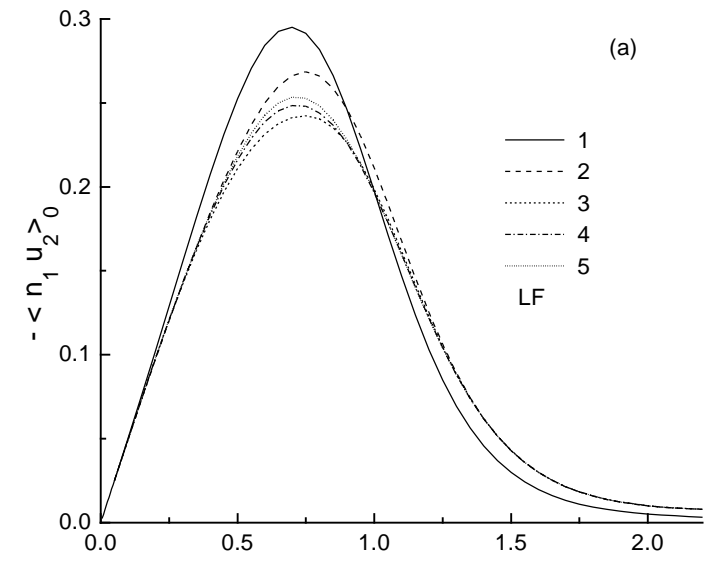
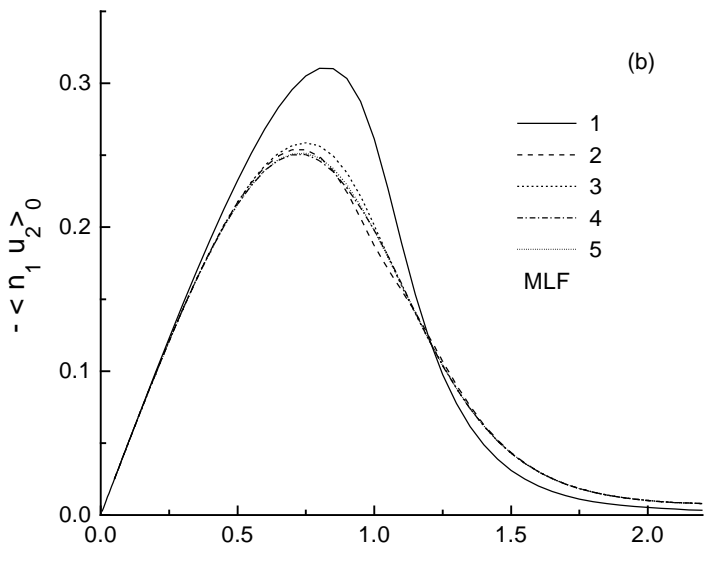
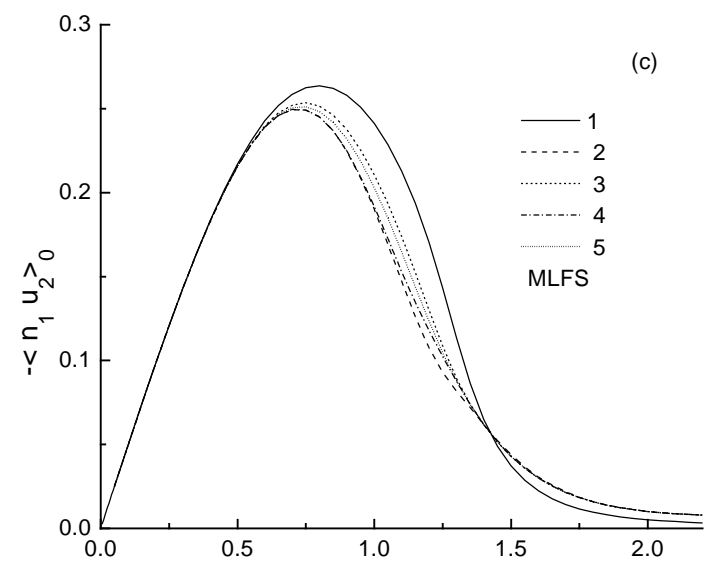
FIG. 3. Plot of the correlation function $\langle n_1u_2 \rangle_{0}$ 
calculated up to different order of perturbations in the wave function 
versus $g_{+}$ for $t/\omega_0 =0.5$ 
within different methods (a) LF, (b) MLF and (c) MLFS. 
The labels (2), (3), .... denote the curve obtained by considering  
up to the second, third, .... order correction to the wave function, 
respectively.
\vskip 0.5cm
\noindent

FIG. 4. Variation of the relative perturbation corrections 
$E_{0}^{(n)}/E_{0}^{(0)}$ to the ground state energy as a 
function of the coupling strength ($g_{+}$) for $t/\omega_0 =1.1$ 
for (a) LF, (b) MLF and (c) MLFS method.
$E_{0}^{(n)}$ is the nth order perturbation correction to the 
ground state energy and $E_{0}^{(0)}$ is the unperturbed ground 
state energy. 
\vskip 0.5 cm
\noindent

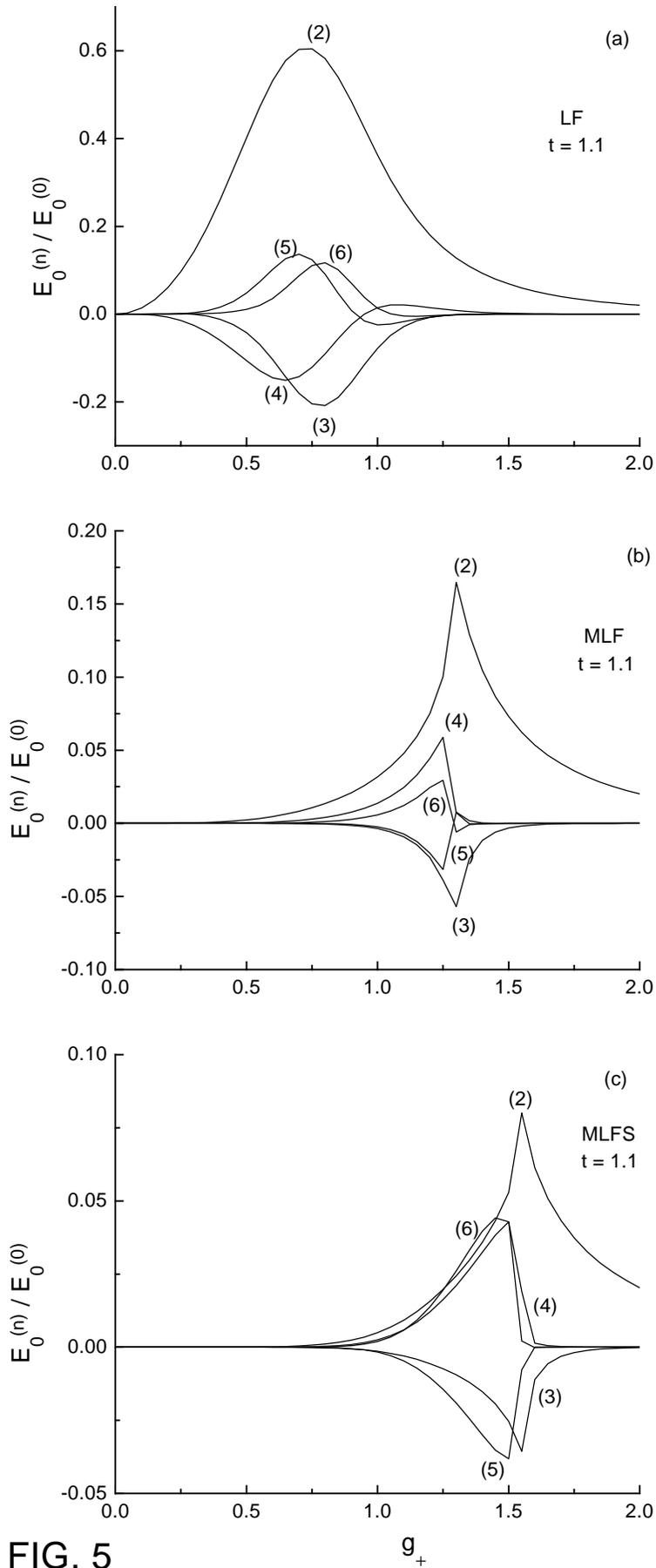
FIG. 5.The plot of the ground state energy obtained by considering 
perturbation correction up to fifth order within the MLF method 
as a function of $g_+$ for $t/\omega_0 =1.1$. 
The unperturbed energies (or with zero phonon averaging results (ZPA)) 
within LF, MLF and MLFS methods are also shown.
\vskip 0.5 cm
\noindent

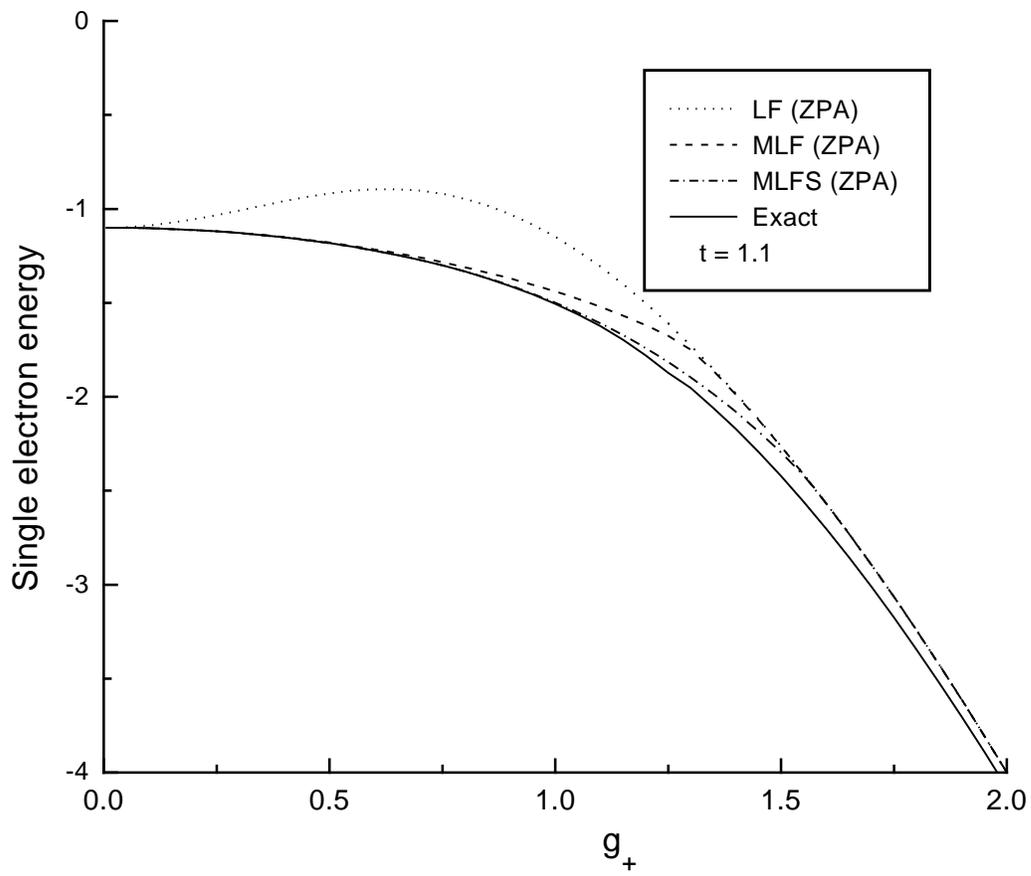
FIG. 6. Plot of the correlation function $\langle n_1u_2 \rangle_{0}$ 
calculated up to different order of perturbations in the wave function 
versus $g_{+}$ for $t/\omega_0 =1.1$ 
within different methods (a) LF, (b) MLF and (c) MLFS. 
The labels (2), (3), .... denote the curve obtained by considering up to  
the second, third,... order correction to the wave function, respectively.
\vskip 0.5cm
\noindent

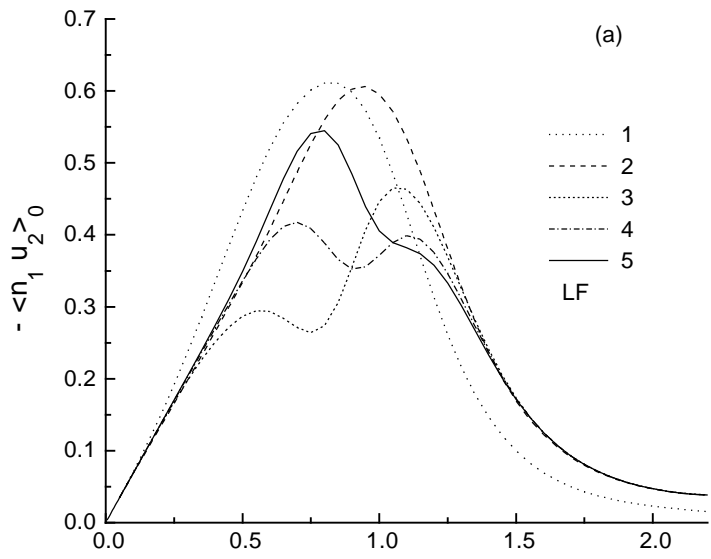
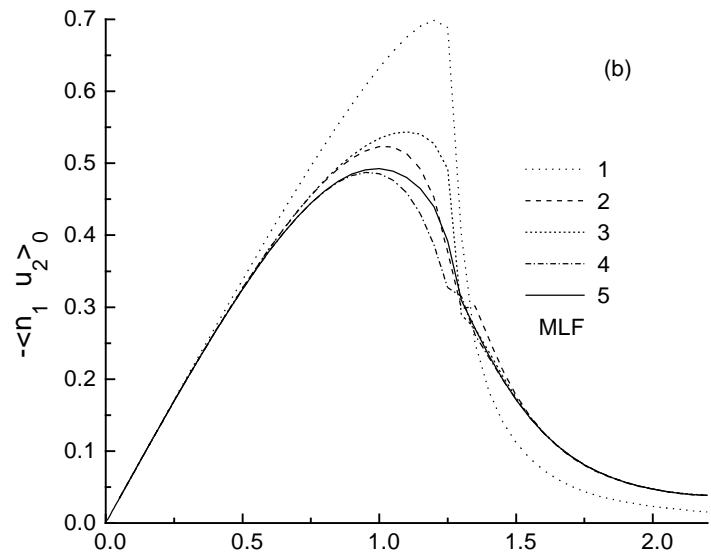
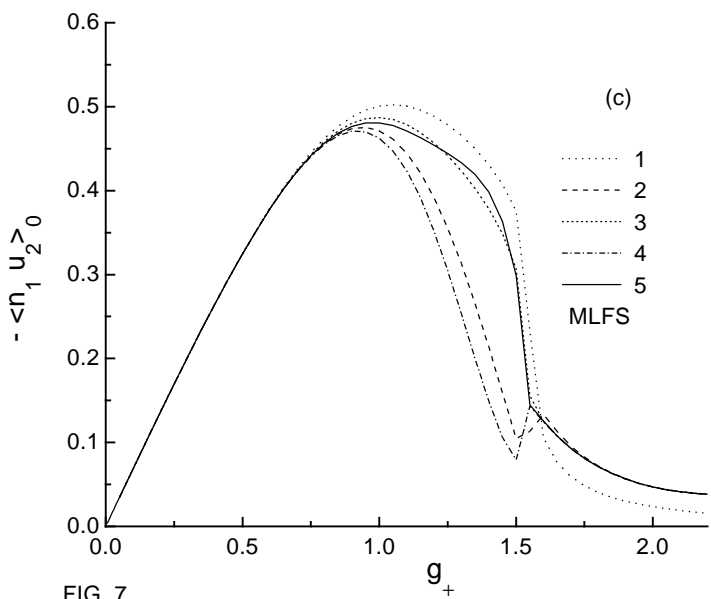
FIG. 7. The variations of $\langle n_{1} (u_{1}-u_{2}) \rangle_{0}/g_+$  
 with $g_+$ for LF, MLF and MLFS methods with zeroth 
order of perturbation for $t/\omega_0=1.1$. 
The solid curve corresponds to that 
obtained within the MLF method up to fifth order correction to wave function.
\vskip 0.5cm
\noindent

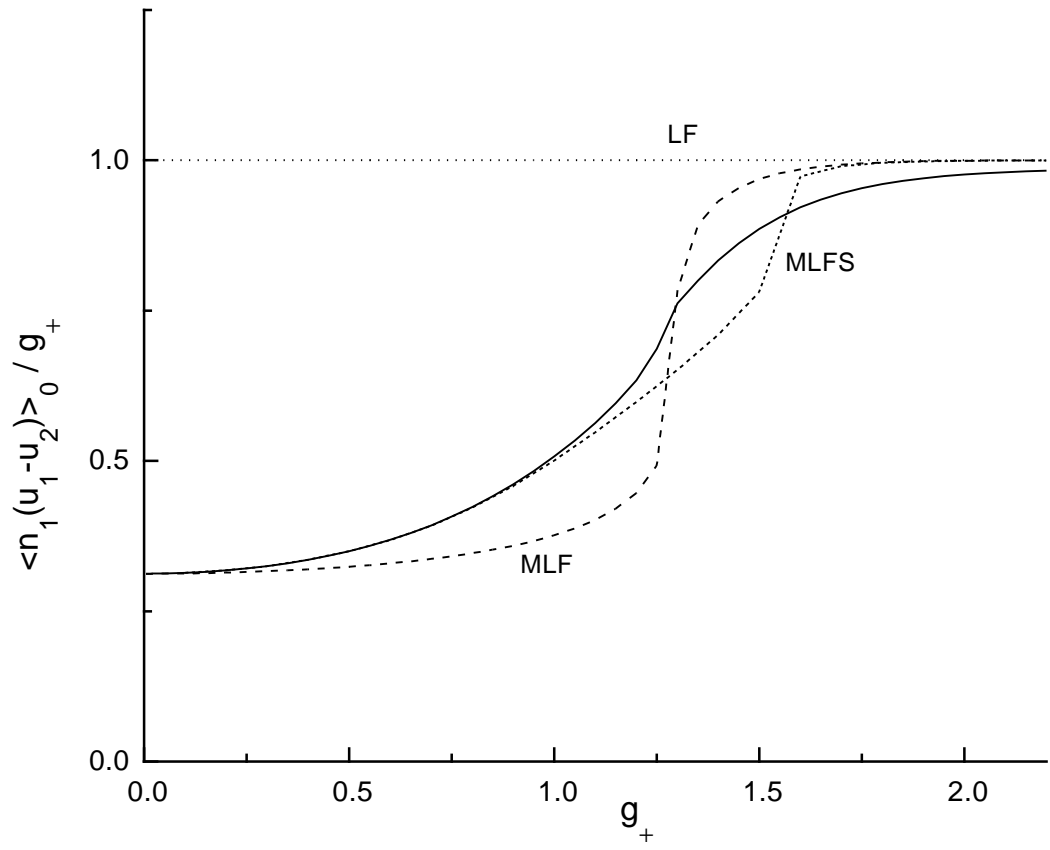
FIG. 8. Ground state oscillator wave function $G(x)$ as a function 
of x for different values of the coupling strength when the electron 
is located at site 1 for $t/\omega_0 =1.1$. 
\vskip 0.5 cm     
\noindent

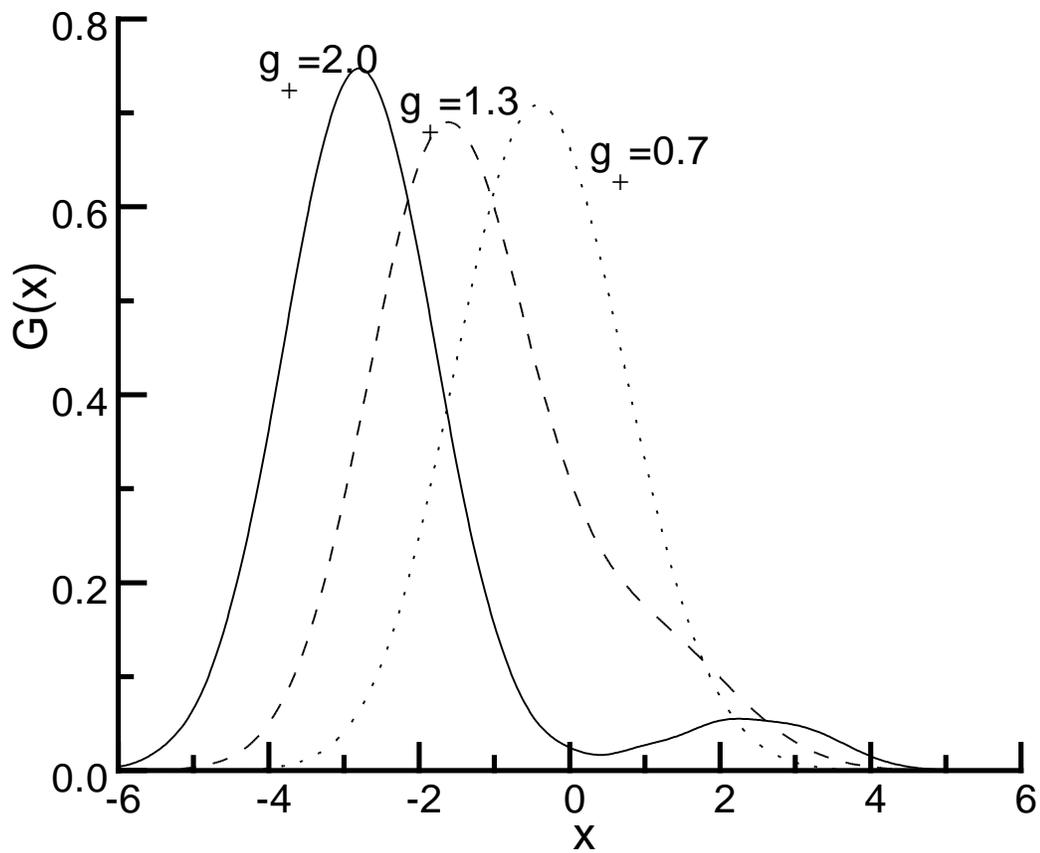
FIG. 9. Variation of the relative perturbation corrections 
$E_{0}^{(n)}/E_{0}^{(0)}$ to the ground state energy as a 
function of the coupling strength ($g_{+}$) for $t/\omega_0 =2.1$ 
for (a) LF, (b) MLF and (c) MLFS method.
$E_{0}^{(n)}$ is the nth order perturbation correction to the 
ground state energy and $E_{0}^{(0)}$ is the unperturbed ground state energy. 

\vskip 0.5 cm
\noindent

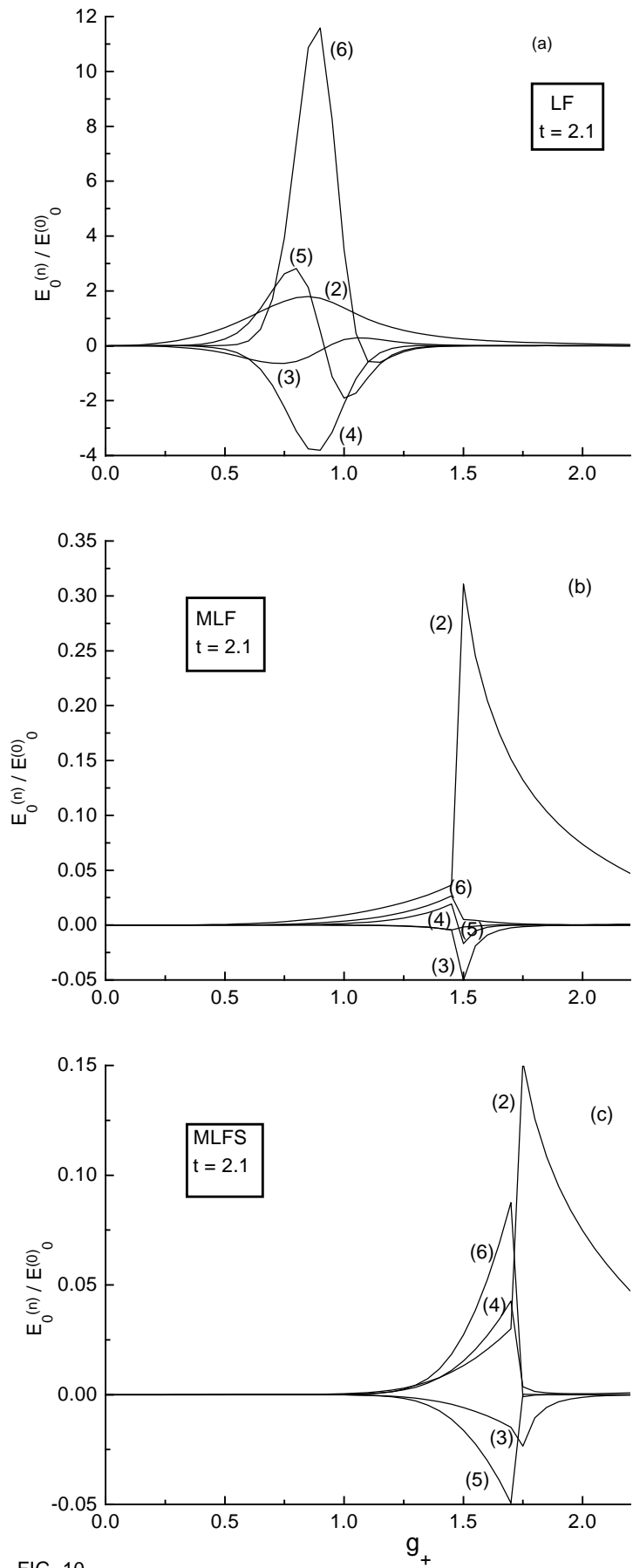
FIG. 10. The variation of $|\langle \psi_0^{(0)}|\psi_G\rangle|^2$ 
with $g_+$ 
within LF, MLF and MLFS methods for $t/\omega_0 =2.1$. $|\psi_G\rangle$ 
and $|\psi_0^{(0)}\rangle$ are the corrected and the unperturbed wave 
functions, respectively.
\vskip 0.5 cm
\noindent

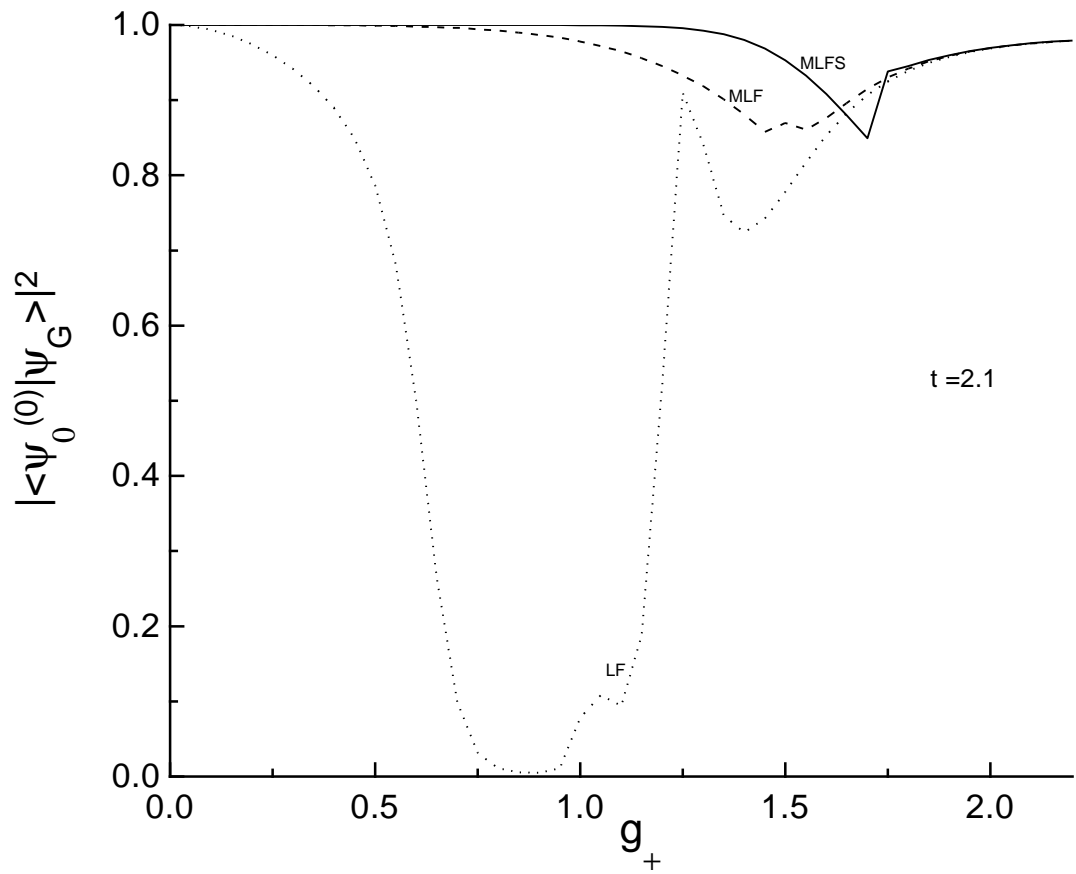
FIG. 11. Plot of the correlation function $\langle n_1u_2 \rangle_{0}$ 
calculated up to different order of perturbations in the wave function 
versus $g_{+}$ for $t/\omega_0 =2.1$ 
within different methods (a) LF, (b) MLF and (c) MLFS. 
The labels (2), (3), .... denote the curve obtained by considering up 
to the  second, third, .... order correction to the wave function, 
respectively.
\vskip 0.5cm
\noindent

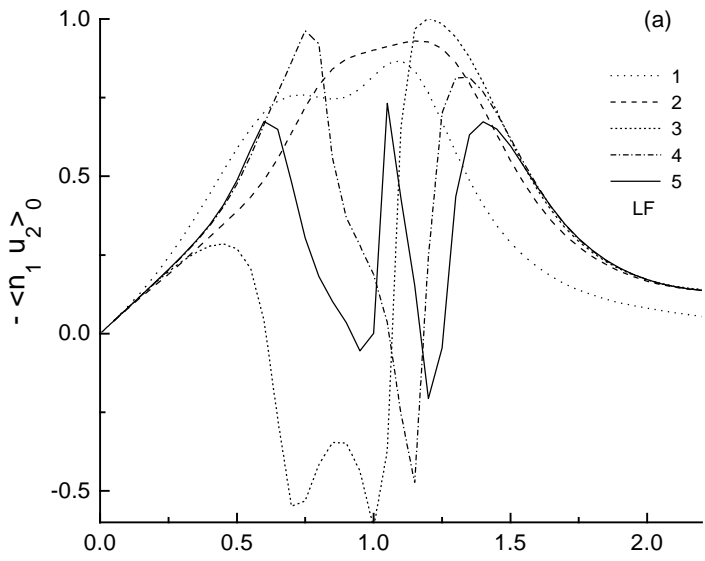
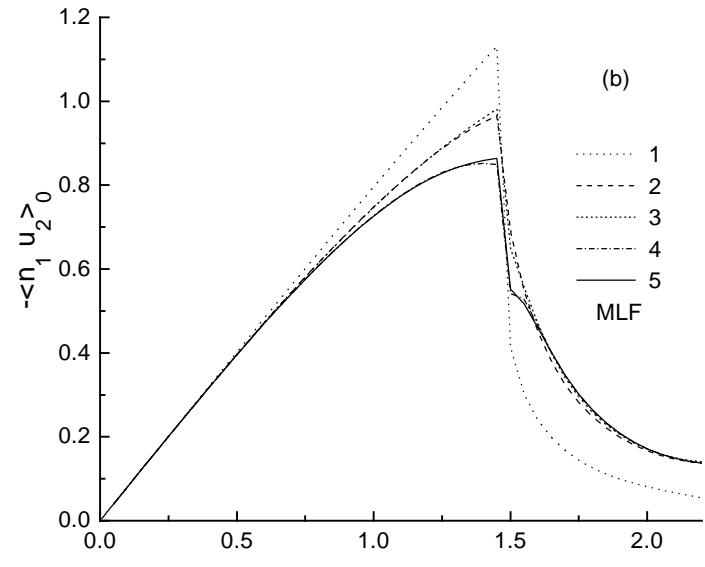
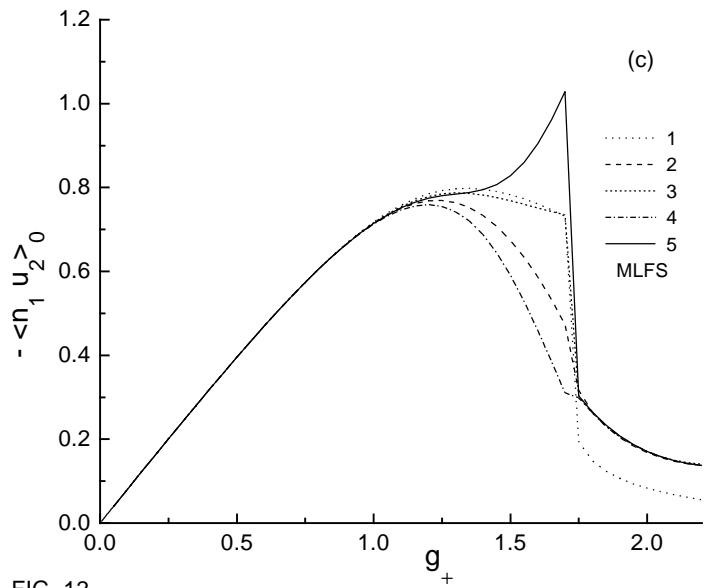
FIG. 12. Ground state oscillator wave function $G(x)$ as a function 
of x for $g_+ =1.6$ when the electron is located at site 1 with 
different values of $t/\omega_0$.
\vskip 0.5 cm     
\noindent

\end{document}